\documentclass[aps,prd,superscriptaddress,twoside,twocolumn,nofootinbib,showpacs,floatfix]{revtex4-1}
\usepackage{amsmath,amssymb}
\usepackage{graphicx,bm,braket}
\usepackage{subfigure}
\pdfpagewidth=\paperwidth
\pdfpageheight=\paperheight
\allowdisplaybreaks
\usepackage{epstopdf}
\newcommand*{\slashed}[1]{{#1\!\!\!/}}
\newcommand*{\hc}{\text{H.\,c.}}

\begin{document}

\title{\boldmath Photoproduction $\gamma p \to f_0(980) p$ in an effective Lagrangian approach}

\author{Neng-Chang Wei}
\affiliation{School of Physics, Henan Normal University, Henan 453007, China}
\affiliation{School of Nuclear Science and Technology, University of Chinese Academy of Sciences, Beijing 101408, China}

\author{Ai-Chao Wang}
\affiliation{College of Science, China University of Petroleum (East China), Qingdao 266580, China}

\author{Fei Huang}
\email{Corresponding author. Email: huangfei@ucas.ac.cn}
\affiliation{School of Nuclear Science and Technology, University of Chinese Academy of Sciences, Beijing 101408, China}

\date{\today}

\begin{abstract}
The most recent data on differential cross sections and photon beam asymmetries $\Sigma$ from the LEPS2/BGOegg Collaboration for the $\gamma p \to f_0(980) p$ reaction are analyzed within a tree-level effective Lagrangian approach. The $t$-channel $\rho$ and $\omega$ exchanges, the $u$- and $s$-channel $N$ exchanges, the interaction current, and the possible $s$-channel $N^*$ exchanges are taken into account in constructing the reaction amplitudes to reproduce the data. The results show that the contributions from either the $N(2040)3/2^+$ or the $N(2100)1/2^+$ resonance exchange in the $s$ channel are necessarily required to describe the LEPS2/BGOegg data and they dominate the differential cross sections of $\gamma p \to f_0(980) p$. Further analysis shows that the contributions from the $t$-channel $\rho$ and $\omega$ exchanges and the interaction current are rather small to both differential cross sections and photo beam asymmetries, and the contributions from $u$-channel $N$ exchange are considerable in the case of including $N(2040)3/2^+$ in the model while negligible in the case of including $N(2100)1/2^+$ in the model. Predictions of target asymmetries $T$ for $\gamma p \to f_0(980) p$ are given, which can be examined by future experiments.
\end{abstract}



\maketitle

\section{Introduction}   \label{Sec:intro}

For a long time, our knowledge of the low mass scalar mesons has been mainly coming from the hadron-induced reactions by using hadronic probes such as $\pi$, $p$ or $\bar{p}$  beams, from the $\gamma \gamma$ collisions, and from the decays of heavier mesons such as $\phi$, $J/\psi$, $D$, and $B$ \cite{ParticleDataGroup:2022pth,CLAS:2008ycy}. However, the situation has changed significantly in the past 20 years as the experimental technology advances in the electromagnetic accelerator facilities, including the control and polarization of photon beams, development of polarized gas and solid targets, construction of large solid angle detectors, development of higher rate data acquisition systems and of data analysis and statistical techniques \cite{Ireland:2019uwn}, have made the electromagnetic probes become an effective and valuable tool for the investigation of conventional and exotic mesons \cite{ Kohri:2009xe, Wieland:2010cq, Moriya:2013hwg, CLAS:2021osv, Mokeev:2022afv, Ireland:2019uwn, Laget:2019tou, Aznauryan:2011qj, CLAS:2017vxx, CLAS:2008ycy, CLAS:2009ngd, CLAS:2018azo,CLAS-beam,Zachariou:2020kkb}.

Stimulated by the most recent data on differential cross sections and photon beam asymmetries $\Sigma$ for the $\gamma p \to f_0(980) p$ reaction with $f_0(980)$ decaying into $\pi^0 \pi^0$ released by the LEPS2/BGOegg Collaboration \cite{LEPS2BGOegg:2023ssr}, in the present work we focus on the analysis of the isoscalar scalar meson $f_0(980)$ photoproduction.

Experimentally, suffering from the dominance of vector meson photoproduction in the data sample, the data of the $f_0(980)$ photoproduction is scarce. Thanks to the high-intensity and high-quality tagged-photon beams produced at The Thomas Jefferson National Accelerator Facility, the CLAS Collaboration has reported the first data on differential cross sections for $\gamma p \to f_0(980) p$ at one photon beam energy point around $E_\gamma \approx 3.4$ GeV with relative low statistics \cite{CLAS:2008ycy, CLAS:2009ngd, CLAS:2018azo}. Since the CLAS detector is optimized for exclusive meson photoproduction with charged mesons in the final states, the CLAS measurements of the $f_0(980)$ photoproduction have been performed by detecting the $f_0(980)$ decaying into $\pi^+ \pi^-$ or $K^+ K^-$. Most recently, the LEPS2/BGOegg Collaboration released the data on differential cross sections and photon beam asymmetries $\Sigma$ for photoproduction of the $f_0(980)$ meson decaying into $\pi^0 \pi^0$ at energies from the reaction threshold up to $E_\gamma \approx 2.4$ GeV  \cite{LEPS2BGOegg:2023ssr}. Since the vector meson cannot decay into $\pi^0 \pi^0$, the LEPS2/BGOegg measurements of the $f_0(980)$ photoproduction are free from the influence of the large contributions from the $\rho$ photoproduction and the so-called S-P interference  \cite{LEPS2BGOegg:2023ssr}. What's more, the LEPS2/BGOegg measurements of the polarization observable $\Sigma$ can provide additional constraints on the phenomenological models for the $f_0(980)$ photoproduction and thus help to understand more reliably the production mechanisms of the $f_0(980)$ meson.

Theoretically, the pioneering CLAS data \cite{CLAS:2008ycy, CLAS:2009ngd, CLAS:2018azo} for $\gamma p \to f_0(980) p$ have been analyzed by several works based on Regge models and/or effective Lagrangian approaches. In Ref.~\cite{daSilva:2013yka}, the $\gamma p \to f_0(980) p$ reaction was analyzed in a Regge model with distinct scenarios for the $f_0(980)\to V\gamma$ decay process being considered. It was shown that the radiative decay rates for $f_0(980)\to V\gamma$ are important in theoretical predictions. In Ref.~\cite{Donnachie:2015jaa}, the $\gamma p \to f_0(980) p$ reaction was analyzed based on a previous Regge-pole model for the $\pi^0$ photoproduction \cite{Barker:1977pm} by further taking into account the Regge cuts, and the authors concluded that the results indicate strongly the presence of the $K\bar{K}$ molecular component in $f_0(980)$. In Ref.~\cite{Lee:2016vlw}, the photoproduction of $f_0(980)$ was analyzed by using an effective Lagrangian approach with the $t$-channel amplitude being constructed by replacing the Feynman propagator with the Regge propagator for $\rho$ exchange to describe the CLAS data \cite{CLAS:2008ycy, CLAS:2009ngd}. In Ref.~\cite{Xing:2018axn}, the $f_0(980)$ photoproduction was analyzed in the reaction threshold region within an effective Lagrangian approach with both the $t$-channel $\rho$ and $\omega$ exchanges being taken into account to reproduce the CLAS data \cite{CLAS:2008ycy, CLAS:2009ngd, CLAS:2018azo}. Besides, the $f_0(980)$ photoproduction has also been investigated in a coupled channel analysis of the $S$-wave $\pi \pi$ and $K\bar{K}$ photoproduction in Ref.~\cite{Ji:1997fb} and in $\pi \pi p$, $\pi \eta p$, and $K\bar{K} p$ channels to examine the $a_0(980)$-$f_0(980)$ mixing in Ref.~\cite{Tarasov:2013yma}.

In the present work, we construct a theoretical model to analyze the newly released LEPS2/BGOegg data for $\gamma p \to f_0(980) p$ \cite{LEPS2BGOegg:2023ssr} based on a tree-level effective Lagrangian approach. The major object is to understand the reaction mechanisms of the $f_0(980)$ photoproduction and to learn the information about possible $N^*$'s exchanged in the $s$ channel of this reaction. Note that, as mentioned above, the LEPS2/BGOegg data cover a much wider energy range, i.e. from threshold up to $E_\gamma \approx 2.4$ GeV, than the CLAS data that are measured at only one energy point around $E_\gamma \approx 3.4$ GeV. Moreover, the LEPS2/BGOegg Collaboration also provides for the first time the data on photon beam asymmetries $\Sigma$ in addition to the differential cross sections. The present work presents so far the first theoretical analysis of the newly reported LEPS2/BGOegg data on both differential cross sections and photon beam asymmetries for $\gamma p \to f_0(980) p$, and is expected to achieve more reliable theoretical results and a better understanding of the reaction mechanisms of $\gamma p \to f_0(980) p$.

The paper is organized as follows. In Sec.~\ref{Sec:formalism}, we introduce the basic formalism of our theoretical model by giving the explicit expressions of the Lagrangians for the interaction vertices, the resonance propagators, and the phenomenological form factors. The model results are shown and discussed in detail in Sec.~\ref{Sec:results}. A brief summary and the conclusions of the present paper are given in Sec.~\ref{sec:summary}.

\section{Formalism}  \label{Sec:formalism}

\begin{figure}[tbp]
\centering
{\vglue 0.15cm}
\subfigure[~$s$ channel]{
\includegraphics[width=0.45\columnwidth]{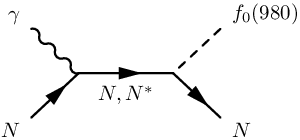}}  {\hglue 0.4cm}
\subfigure[~$t$ channel]{
\includegraphics[width=0.45\columnwidth]{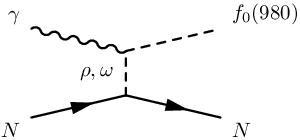}} \\[6pt]
\subfigure[~$u$ channel]{
\includegraphics[width=0.45\columnwidth]{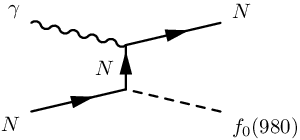}} {\hglue 0.4cm}
\subfigure[~Interaction current]{
\includegraphics[width=0.45\columnwidth]{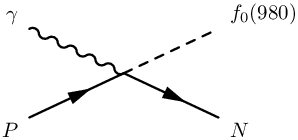}}
\caption{Generic structure of the amplitude for $\gamma N \to f_0(980) N$. Time proceeds from left to right.}
\label{FIG:feymans}
\end{figure}

For the convenience of discussion, we express the reaction of interest as follow
\begin{equation}
\gamma(k) + p(p) \to f_0(980)(q) + p(p^\prime).
\label{eq:reaction}
\end{equation}
Here, the arguments $k$, $p$, $q$, and $p^\prime$ in parentheses stand for the corresponding four momenta of the initial and final particles. The Mandelstam variables $s$, $t$, and $u$ are defined as usual: $t \equiv (p - p^\prime)^2 = (k - q)^2$, $s \equiv (p + k)^2 = (q + p^\prime)^2$, $u \equiv (p - q)^2 = (p^\prime - k)^2$.

In the present paper, we perform an analysis of the most recent LEPS2/BGOegg data for the $\gamma p \to f_0(980) p$ reaction within a tree-level effective Lagrangian approach. As diagrammatically depicted in Fig.~\ref{FIG:feymans}, we consider the following reaction mechanisms to construct the reaction amplitudes for $\gamma p \to f_0(980) p$: (a) the $s$-channel $N$ and $N^*$exchanges, (b) the $t$-channel $\rho$ and $\omega$ exchanges, (c) the $u$-channel $N$ exchange, and (d) the interaction current. In principle the $u$-channel $N^\ast$ exchange should also be considered, which will cause an additional parameter to the model, i.e. the cutoff parameter in the form factor. We have checked and found that the contributions from the $u$-channel $N^\ast$ exchange are rather small and have negligible effects on both fit qualities and the values of extracted parameters. Thus, this dynamics is not included in the present work.  According to Fig.~\ref{FIG:feymans}, the transition current $M^{\mu}$ can be written as \cite{Haberzettl:1997,Haberzettl:2006bn,Huang:2012,Huang:2012xj}
\begin{equation}
M^{\mu} \equiv M^{\mu}_s + M^{\mu}_t + M^{\mu}_u + M^{\mu}_{\rm int},  \label{eq:amplitude}
\end{equation}
with $\mu$ denoting the Lorentz index of the photon field. $M^{\mu}_s$, $M^{\mu}_t$, and $M^{\mu}_u$ represent the transition currents calculated from the $s$-, $t$-, and $u$-channel diagrams, respectively. The explicit expressions of $M^{\mu}_s$, $M^{\mu}_t$, and $M^{\mu}_u$ can be straightforwardly obtained from the effective Lagrangians, propagators, and form factors given below. The last term in Eq.~(\ref{eq:amplitude}), $M^{\mu}_{\rm int}$, is the interaction current term, which stands for the contributions calculated from the diagrams that do not have $s$-, $t$-, or $u$-channel poles and is responsible for preserving gauge invariance of the full reaction amplitudes. In the present work, we follow Refs.~\cite{Haberzettl:2006bn,Huang:2012xj} to model the interaction current $M^{\mu}_{\rm int}$ by a generalized contact current
\begin{equation}
M^{\mu}_{\rm int} = \Gamma_{NNf_0(980)}(q) C^\mu,
\label{eq:Mint}
\end{equation}
with $\Gamma_{NNf_0(980)}(q)$ standing for the vertex function of the $NNf_0(980)$ interaction obtained from the Lagrangian of Eq.~(\ref{eq:nnf0}). The auxiliary current $C^\mu$ is introduced to ensure that the total reaction amplitude of $\gamma p \to f_0(980) p$ satisfies the generalized Ward-Takahashi identity and thus is fully gauge invariant. Following Refs.~\cite{Haberzettl:1997,Haberzettl:2006bn,Huang:2012,Huang:2012xj}, the prescription for $C^\mu$ for the $\gamma p \to f_0(980) p$ reaction can be written as
\begin{equation}
C^\mu = - Q_{u} \frac{f_{u}-\hat{F}}{u-p^{\prime 2}} (2p^{\prime}-k)^\mu - Q_{s} \frac{f_{s}-\hat{F}}{s-p^2} (2p+k)^\mu,
\label{eq:c}
\end{equation}
with $Q_u$ and $Q_s$ being the charges of the nucleon exchanged in the $u$ channel and $s$ channel, respectively. The expression of $\hat{F}$ in Eq.~(\ref{eq:c}) is
\begin{equation}
\hat{F} = 1 - \hat{h} \left(1 -  f_u\right) \left(1 -  f_s\right),
\end{equation}
where $f_u$ and $f_s$ are the phenomenological form factors attaching to the amplitudes of the $N$ exchanged in the $u$ and $s$ channels, respectively. $\hat{h}$ is set to be $1$ for simplicity as usual \cite{Wang:2017tpe,Wang:2018vlv}.

\subsection{Effective Lagrangians} \label{Sec:Lagrangians}

The explicit expressions of the effective Lagrangians for the interaction vertices are given in this subsection. For the sake of simplicity, the notation $f_0$ is used to stand for $f_0(980)$ and we define the following operators
\begin{equation}
\Gamma^{(+)}=\gamma_5 \qquad {\rm and} \qquad \Gamma^{(-)}=1,
\end{equation}
and the field-strength tensor of the photon field $A^\mu$
\begin{equation}
F^{\mu\nu} = \partial^\mu A^\nu - \partial^\nu A^\mu.
\end{equation}

To calculate the amplitudes of the $s$- and $u$-channel $N$ exchanges, we use the following Lagrangians
\begin{eqnarray}
{\cal L}_{\gamma NN} &=& -\,e \bar{N} \!\left[ \! \left( \hat{e} \gamma^\mu - \frac{ \hat{\kappa}_N} {2M_N}\sigma^{\mu \nu}\partial_\nu \! \right) \! A_\mu\right]\! N,  \\[6pt]
{\cal L}_{NNf_0} &=& g_{NNf_0}\bar{N}Nf_0,               \label{eq:nnf0}
\end{eqnarray}
with $e$ being the elementary charge unit and $\hat{e}$ representing the charge operator acting on the $N$ field. $\hat{\kappa}_N \equiv \kappa_p\hat{e} + \kappa_n(1-\hat{e})$ is the anomalous magnetic moment with $\kappa_p=1.793$ for proton and $\kappa_n=-1.913$ for neutron. Since both the experimental and theoretical information on the coupling constant $g_{NNf_0}$ are scarce, its value will be determined by adjusting to reproduce the available data for $\gamma p \to f_0(980) p$.

The following Lagrangians are employed to calculate the amplitudes of the $t$-channel $\rho$ and $\omega$ exchanges
\begin{eqnarray}
{\cal L}_{\omega NN} &=& -\,g_{\omega NN} \bar{N} \!\left( \gamma^\mu - \frac{\kappa_\omega}{2M_N}\sigma^{\mu \nu}\partial_\nu  \! \right) \! \omega_\mu N,      \\[6pt]
{\cal L}_{\gamma f_0 \omega} &=& \frac{eg_{\gamma f_0 \omega}}{2M_\omega} \partial_\mu A_\nu \!\left( \partial^\mu\omega^\nu - \partial^\nu\omega^\mu \! \right) \! f_0,        \\[6pt]
{\cal L}_{\rho NN} &=& -\,g_{\rho NN} \bar{N} \!\left( \gamma^\mu - \frac{\kappa_\rho}{2M_N}\sigma^{\mu \nu}\partial_\nu  \! \right) \! \vec{\tau} \cdot \vec{\rho}_\mu N,  \\[6pt]
{\cal L}_{\gamma f_0 \rho} &=& \frac{eg_{\gamma f_0 \rho}}{2M_\rho} \partial_\mu A_\nu \!\left( \partial^\mu\rho^\nu - \partial^\nu\rho^\mu \! \right) \! f_0.
\end{eqnarray}
Here, the coupling constants $g_{\omega NN} = 15.85$, $\kappa_\omega = 0$, $g_{\rho NN} = 3.36$, and $\kappa_\rho = 6.1$ are taken from Ref.~\cite{Xing:2018axn}. The constants $g_{\gamma f_0 \omega} = 0.58$ and $g_{\gamma f_0 \rho} = 0.61$ are calculated from the partial decay widthes of $\Gamma_{f_0 \to \omega\gamma} = 6.6$ keV and $\Gamma_{f_0 \to \rho\gamma} = 7.3$ keV obtained in Ref.~\cite{Nagahiro:2008mn}.

To calculate the amplitudes of the $s$-channel $N^*$ exchanges, we use the following Lagrangians for the electromagnetic interaction vertices
\begin{eqnarray}
{\cal L}_{RN\gamma}^{1/2\pm} &=& e\frac{g_{RN\gamma}^{(1)}}{2M_N}\bar{R} \Gamma^{(\mp)}\sigma_{\mu\nu} \left(\partial^\nu A^\mu \right) N  + \hc,  \label{eq:r1} \\[6pt]
{\cal L}_{RN\gamma}^{3/2\pm} &=& -\, ie\frac{g_{RN\gamma}^{(1)}}{2M_N}\bar{R}_\mu \gamma_\nu \Gamma^{(\pm)}F^{\mu\nu}N \nonumber \\
&&+\, e\frac{g_{RN\gamma}^{(2)}}{\left(2M_N\right)^2}\bar{R}_\mu \Gamma^{(\pm)}F^{\mu \nu}\partial_\nu N + \hc, \\[6pt]
{\cal L}_{RN\gamma}^{5/2\pm} & = & e\frac{g_{RN\gamma}^{(1)}}{\left(2M_N\right)^2}\bar{R}_{\mu \alpha}\gamma_\nu \Gamma^{(\mp)}\left(\partial^{\alpha} F^{\mu \nu}\right)N \nonumber \\
&& \pm\, ie\frac{g_{RN\gamma}^{(2)}}{\left(2M_N\right)^3}\bar{R}_{\mu \alpha} \Gamma^{(\mp)}\left(\partial^\alpha F^{\mu \nu}\right)\partial_\nu N \nonumber \\
&& + \,  \hc,  \\[6pt]
{\cal L}_{RN\gamma}^{7/2\pm} &=&  ie\frac{g_{RN\gamma}^{(1)}}{\left(2M_N\right)^3}\bar{R}_{\mu \alpha \beta}\gamma_\nu \Gamma^{(\pm)}\left(\partial^{\alpha}\partial^{\beta} F^{\mu \nu}\right)N \nonumber \\
&&-\, e\frac{g_{RN\gamma}^{(2)}}{\left(2M_N\right)^4}\bar{R}_{\mu \alpha \beta} \Gamma^{(\pm)} \left(\partial^\alpha \partial^\beta F^{\mu \nu}\right) \partial_\nu N  \nonumber \\
&&  + \,  \hc,
\end{eqnarray}
and use the following Lagrangians for the hadronic interaction vertices
\begin{eqnarray}
{\cal L}_{RNf_0}^{1/2\pm} &=& -i g_{RNf_0} \bar{N}\Gamma^{(\mp)} f_0 R +\, \hc, \label{eq:half1}    \\[6pt]
{\cal L}_{RNf_0}^{3/2\pm} &=& \mp \frac{g_{RNf_0}}{M_{f_0}} \bar{N}\Gamma^{(\pm)} \left(\partial^{\alpha}f_0\right) R_{\alpha} +\hc,      \\[6pt]
{\cal L}_{RNf_0}^{5/2\pm} &=& i \frac{g_{RNf_0}}{M_{f_0}^2} \bar{N}\Gamma^{(\mp)} \left(\partial^{\alpha}\partial^{\beta}f_0\right) R_{\alpha\beta} +\hc,      \\[6pt]
{\cal L}_{RNf_0}^{7/2\pm} &=& \pm\frac{g_{RNf_0}}{M_{f_0}^3} \bar{N}\Gamma^{(\pm)}\left(\partial^{\alpha}\partial^{\beta}\partial^{\gamma}f_0\right)R_{\alpha\beta\gamma}  \nonumber \\
      && + \, \hc.  \label{eq:r8}
\end{eqnarray}
In Eq.~(\ref{eq:r1})-(\ref{eq:r8}), $R$ represents the $N^*$ resonances and the superscript of ${\cal L}$ denotes the spin and parity of the resonance $R$. In our tree-level calculation in the present paper, the model results are only sensitive to the products of hadronic coupling constant $g_{RNf_0}$ and the electromagnetic coupling constants $g_{RN\gamma}^{(i)}$ ($i=1,2$). If the helicity amplitudes of $R\to N\gamma$ are available in Review of Particle Physics (RPP) \cite{ParticleDataGroup:2022pth}, they will be used to calculate the corresponding electromagnetic coupling constants. Otherwise, the products of the hadronic coupling constant and electromagnetic coupling constants will be determined by fitting to the available data of $\gamma p \to f_0(980) p$.

\subsection{Resonance propagators}

By defining
\begin{eqnarray}
\tilde{g}_{\mu \nu} &=& -\, g_{\mu \nu} + \frac{p_{\mu} p_{\nu}}{M_R^2}, \\[6pt]
\tilde{\gamma}_{\mu} &=& -\, \gamma_{\mu} + \frac{p_{\mu}\slashed{p}}{M_R^2},   \label{eq:prop-auxi}
\end{eqnarray}
the propagators of the resonances with spin $1/2$, $3/2$, $5/2$, and $7/2$ employed in the present work adopt the following prescriptions \cite{Wang:2017tpe}:
\begin{eqnarray}
S_{1/2}(p) &=& \frac{i}{\slashed{p} - M_R + i \Gamma_R/2}, \label{propagator-1hf}  \\[6pt]
S_{3/2}(p) &=&  \frac{i}{\slashed{p} - M_R + i \Gamma_R/2} \left( \tilde{g}_{\mu \nu} + \frac{1}{3} \tilde{\gamma}_\mu \tilde{\gamma}_\nu \right),  \label{propagator-3hf} \\[6pt]
S_{5/2}(p) &=&  \frac{i}{\slashed{p} - M_R + i \Gamma_R/2} \,\bigg[ \, \frac{1}{2} \big(\tilde{g}_{\mu \alpha} \tilde{g}_{\nu \beta} + \tilde{g}_{\mu \beta} \tilde{g}_{\nu \alpha} \big)  \nonumber \\
&& -\, \frac{1}{5}\tilde{g}_{\mu \nu}\tilde{g}_{\alpha \beta}  + \frac{1}{10} \big(\tilde{g}_{\mu \alpha}\tilde{\gamma}_{\nu} \tilde{\gamma}_{\beta} + \tilde{g}_{\mu \beta}\tilde{\gamma}_{\nu} \tilde{\gamma}_{\alpha}  \nonumber \\
&& +\, \tilde{g}_{\nu \alpha}\tilde{\gamma}_{\mu} \tilde{\gamma}_{\beta} +\tilde{g}_{\nu \beta}\tilde{\gamma}_{\mu} \tilde{\gamma}_{\alpha} \big) \bigg], \\[6pt]
S_{7/2}(p) &=&  \frac{i}{\slashed{p} - M_R + i \Gamma_R/2} \, \frac{1}{36}\sum_{P_{\mu} P_{\nu}} \bigg( \tilde{g}_{\mu_1 \nu_1}\tilde{g}_{\mu_2 \nu_2}\tilde{g}_{\mu_3 \nu_3} \nonumber \\
&& -\, \frac{3}{7}\tilde{g}_{\mu_1 \mu_2}\tilde{g}_{\nu_1 \nu_2}\tilde{g}_{\mu_3 \nu_3} + \frac{3}{7}\tilde{\gamma}_{\mu_1} \tilde{\gamma}_{\nu_1} \tilde{g}_{\mu_2 \nu_2}\tilde{g}_{\mu_3 \nu_3} \nonumber \\
&& -\, \frac{3}{35}\tilde{\gamma}_{\mu_1} \tilde{\gamma}_{\nu_1} \tilde{g}_{\mu_2 \mu_3}\tilde{g}_{\nu_2 \nu_3} \bigg),  \label{propagator-7hf}
\end{eqnarray}
where $M_R$ and $\Gamma_R$ denote the mass and width for the resonance $R$ with four-momentum $p$, respectively. The summation over $P_\mu$ $\left(P_\nu\right)$ in Eq.~(\ref{propagator-7hf}) goes over all the $3!=6$ possible permutations of the indices $\mu_1\mu_2\mu_3$ $\left(\nu_1\nu_2\nu_3\right)$. 

We mention that the Rarita-Schwinger prescriptions of resonance propagators, i.e. Eqs.~(\ref{propagator-3hf})-(\ref{propagator-7hf}), have a problem of consisting unphysical components with lower spins, as discussed in Refs.~\cite{Benmerrouche:1989uc,Vrancx:2011qv,Skoupil:2016ast,Mart:2019jtb}. This issue might be resolved if the interaction vertices could be constructed in a proper form with a consistent interaction. A pure spin-$3/2$ propagator with consistent interaction Lagrangians was investigated in Ref.~\cite{Mart:2019jtb}, where it was shown that for the total cross sections of $\pi N$ photoproduction, the pure spin-$3/2$ propagator with consistent interaction Lagrangians and the prescription of Rarita-Schwinger yield similar structures for $s$-channel $\Delta$ contribution. In the present work, the Rarita-Schwinger prescription is employed for resonance propagators as an economic and convenient approximation. A serious treatment of the propagators of resonances with high spins as well as the associated consistent interactions will be done in our future work when more data for this reaction becomes available.

\subsection{Form factors}

In actual calculations, a phenomenological form factor is introduced at each hadronic vertex. In accordance with Refs.~\cite{Huang:2012xj,Wang:2017tpe,Wang:2018vlv}, for the $t$-channel meson exchanges we employ the following form factor
\begin{eqnarray}
f_M(q^2_M) =  \left (\frac{\Lambda_M^2-M_M^2}{\Lambda_M^2-q^2_M} \right)^2,   \label{eq:ff_M}
\end{eqnarray}
with $\Lambda_M$ being the cut off parameter for the $t$-channel meson exchange diagrams, $M_M$ and $q_M$ denoting the mass and four-momentum of the intermediate meson, respectively. For the $s$- and $u$-channel baryon exchanges, the form factor employed in the present paper are \cite{Huang:2012xj,Wang:2017tpe,Wang:2018vlv}
\begin{eqnarray}
f_B(p^2_x) =\left (\frac{\Lambda_B^4}{\Lambda_B^4+\left(p_x^2-M_B^2\right)^2} \right )^2,  \label{eq:ff_B}
\end{eqnarray}
with $\Lambda_B$ being the cut off parameter for $s$- and $u$-channel baryon exchange diagrams, $M_B$ and $p_x$ ($x=s, u$) denoting the mass and four-momentum of the exchanged baryon in $s$ or $u$ channel, respectively. In the present paper, the values of $\Lambda_M$ and $\Lambda_B$ will be determined by the available data for $\gamma p \to f_0(980) p$.

\subsection{Differential cross section}

The CLAS and LEPS2/BGOegg Collaborations have measured the cross sections for the $\gamma p \to f_0(980) p$ reaction with $f_0(980)$ decaying into the $K\bar{K}$ or $\pi\pi$ channels. The signal cross section of $\gamma p \to f_0(980) p$ with $f_0(980)$ decaying into a certain final state $i$ can be calculated via \cite{Donnachie:2015jaa}
\begin{equation}
\frac{{\rm d}\sigma}{{\rm d}\Omega \, {\rm d}M} = \frac{{\rm d}\sigma_0(M)}{{\rm d}\Omega} \frac{2M^2_{f_0}}{\pi} \frac{\Gamma_i(M)}{\left(M_{f_0}^2-M^2\right)^2+M^2\Gamma^2_{f_0}}. 
\label{eq:dif}
\end{equation}
Here, ${\rm d}\sigma_0(M)/{\rm d}\Omega$ is the so-called narrow-width differential cross section at a scalar mass $M$. For a given  value of  $M$, ${\rm d}\sigma_0(M)/{\rm d}\Omega$ can be calculated as the on-shell cross section for $\gamma p \to f_0(980) p$ with a on-shell mass $M$ for $f_0(980)$. In Eq.~(\ref{eq:dif}), $M_{f_0} = 990$ MeV is the mass of $f_0(980)$ and $\Gamma_i$ is the partial decay width for $f_0(980) \to KK$ or $f_0(980) \to \pi \pi$. $M$ is the invariant mass of $KK$ or $\pi \pi$ which is measured in experiments.

\section{Results and discussion}   \label{Sec:results}

The CLAS data on differential cross sections for $\gamma p \to f_0(980) p$ \cite{CLAS:2008ycy, CLAS:2009ngd, CLAS:2018azo} were conducted at photon beam energy $E_\gamma \approx 3.4$ GeV that corresponds to center-of-mass energy $W \approx 2.7$ GeV. In Ref.~\cite{Xing:2018axn}, these data were analyzed within an effective Lagrangian model. It was shown that the CLAS data can be reproduced by considering only the $t$-channel $\rho$ and $\omega$ exchanges with a cutoff value $\Lambda_M=1070$ MeV. 

\begin{figure}[htb]
\includegraphics[width=\columnwidth]{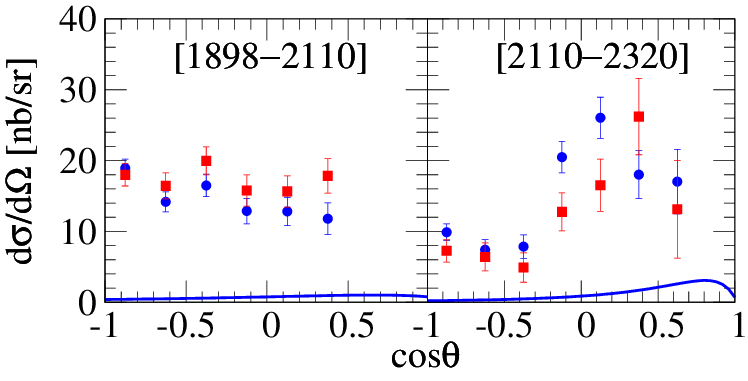}
\caption{Differential cross sections for $\gamma p \to f_0(980) p$ with $f_0(980)$ decaying into $\pi^0 \pi^0$ obtained by considering only the $t$-channel amplitudes of the $\rho$ and $\omega$ exchanges ($\Lambda_M = 1070$ MeV) \cite{Xing:2018axn}. The quantities in square brackets are the ranges of the center-of-mass energy of the system, in MeV. Data are taken from the LEPS2/BGOegg Collaboration \cite{LEPS2BGOegg:2023ssr} for the two event-selection conditions with (squares) and without (circles) requirement of the $N^*$ cuts.}
\label{fig:xie}
\end{figure}

The LEPS2/BGOegg data on both differential cross sections and photon beam asymmetries for $\gamma p \to f_0(980) p$ \cite{LEPS2BGOegg:2023ssr} were measured from threshold up to photon beam energy $E_\gamma \approx 2.4$ GeV ($W \approx 2.3$ GeV). The contributions from $t$-channel $\rho$ and $\omega$ exchanges are rather small in this energy region, as shown in Fig.~\ref{fig:xie}. One sees from this figure that the data are much underestimated if only the $t$-channel interaction is considered, indicating that the contributions from other reaction mechanisms are still needed. We have checked and found that, even if we further included the contributions from the $u$- and $s$-channel $N$ exchanges and the interaction current, the model is still far from being sufficient to describe the data. In this situation, we further considered in the model the possible contributions from $N^*$ exchanges in the $s$ channel.

\begin{table*}[htb]
\caption{$\chi^2/N$ evaluated for the inclusion of one nucleon resonance in the $s$ channel.}
\begin{tabular*}{\textwidth}{@{\extracolsep\fill}ccccccccc}
\hline \hline
$N(2000)5/2^+$& $N(1990)7/2^+$& $N(2040)3/2^+$& $N(2060)5/2^-$& $N(2100)1/2^+$& $N(2120)3/2^-$& $N(2190)7/2^-$& $N(2300)1/2^+$ & Born  \\ \hline
$7.08$& $5.61$& $2.82$& $5.59$& $3.02$& $5.01$& $12.82$& $3.88$ & $14.40$  \\ \hline \hline
\end{tabular*}
\label{table:chis}
\end{table*}

In RPP \cite{ParticleDataGroup:2022pth}, there are seven nucleon resonances with spin $J \leq 7/2$ lying above the reaction threshold of $\gamma p \to f_0(980) p$, namely, the $N(2000)5/2^+$, $N(1990)7/2^+$, $N(2040)3/2^+$, $N(2060)5/2^-$, $N(2100)1/2^+$, $N(2120)3/2^-$, $N(2190)7/2^-$, and $N(2300)1/2^+$ resonances. 
Since we have no clear reason to discard one resonance in favor of the other, we treat these resonances equally by including them one by one in the $s$ channel to reproduce the data. The information on the helicity amplitudes of these resonances is scarce, and there is no information for their decays into $f_0(980) p$ channel either. Thus we treat these quantities as parameters in fitting procedure. For the decay ratio of $f_0(980) \to \pi \pi$, we take the average value of $0.52$, $0.75$, and $0.84$ obtained in Refs.~\cite{BaBar:2006hyf,BES:2005iaq,Anisovich:2001ay} as cited in RPP \cite{ParticleDataGroup:2022pth}. The resulting $\chi^2$ per degree of freedom, $\chi^2/N$, with $N = 38$, for the inclusion of each of those seven nucleon resonances are listed in Table~\ref{table:chis}.

The last column in Table~\ref{table:chis} represents the $\chi^2/N$ evaluated from the Born term where no resonance exchange in the $s$ channel is included in the model. One can see that, the inclusion of $N^*$ in the model can significantly improve the theoretical description of the LEPS2/BGOegg data for $\gamma p \to f_0(980) p$. The three best results are obtained by including the $N(2040)3/2^+$, $N(2100)1/2^+$, and $N(2300)1/2^+$ resonances with $\chi^2/N = 2.82$, $3.02$, and $3.88$, respectively. The fits of including other resonances have much larger $\chi^2/N$ and the corresponding results are not in agreement with the data. These fits are thus not considered as acceptable.

\begin{figure}[htb]
\includegraphics[width=\columnwidth]{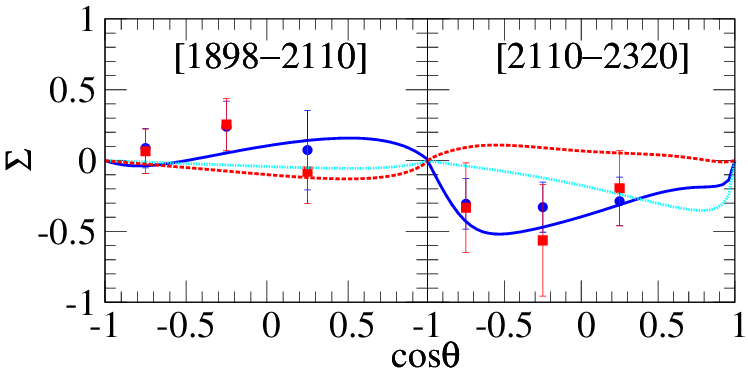}
\caption{ Photon beam asymmetries $\Sigma$ for $\gamma p \to f_0(980) p$ obtained by including one of the $N(2040)3/2^+$ (blue solid lines), $N(2100)1/2^+$ (cyan dotted lines), and $N(2300)1/2^+$ (red dashed lines) resonances. The quantities in square brackets are the ranges of the center-of-mass energy of the system, in MeV. Data are taken from the LEPS2/BGOegg Collaboration \cite{LEPS2BGOegg:2023ssr}  for the two event-selection conditions with (squares) and without (circles) requirement of the $N^*$ cuts.}
\label{fig:compare}
\end{figure}

\begin{table}[htbp]
\caption{Fitted values of model parameters. The asterisks below the resonance names denote the overall rating of each resonance evaluated by RPP \cite{ParticleDataGroup:2022pth}. The numbers in brackets below the resonance mass and width represent the corresponding values estimated by RPP \cite{ParticleDataGroup:2022pth}.}
\label{table:constants}
\renewcommand{\arraystretch}{1.2}
\begin{tabular*}{\columnwidth}{@{\extracolsep\fill}lcc}
\hline\hline
                           & Model I                 & Model II		\\ \hline
$g_{NNf_0}$                     &$-2.08 \pm 0.50$  &$0.65 \pm 0.35$ \\
$\Lambda_B$ $[{\rm MeV}]$       &$958 \pm 11$	   &$800 \pm 15$ \\ \hline
$N^*$Name                  		&$N(2040){3/2}^+$  &$N(2100){1/2}^+$ 	\\
		                        &  $\ast$          & $\ast\ast\ast$ \\
$M_R$ $[{\rm MeV}]$             &  $2038 \pm 1$	   &  $2088 \pm 2$ \\
		                        &                  & [$2050 - 2150$] \\
$\Gamma_R$ $[{\rm MeV}]$        &  $350 \pm 15$	   &  $320 \pm 56$ \\
				                &                  & [$200 - 320$]	\\
$g_{RNf_0}g_{RN\gamma}^{(1)}$   & $-8.83 \pm 0.09$ & $0.53 \pm 0.05$ \\
$g_{RNf_0}g_{RN\gamma}^{(2)}$   & $-5.60 \pm 0.11$ &   \\
\hline\hline
\end{tabular*}
\end{table}

In Fig.~\ref{fig:compare}, we show the comparison of our results of photon beam asymmetries for $\gamma p \to f_0(980) p$ obtained with the inclusion of each of the $N(2040)3/2^+$, $N(2100)1/2^+$, and $N(2300)1/2^+$ resonances. One can see that there are noticeable discrepancies between the results obtained by inclusion of the $N(2300)1/2^+$ resonance with the data. Therefore, we will discuss only the results obtained by including the $N(2040)3/2^+$ or $N(2100)1/2^+$ resonance. We denote the model with $N(2040)3/2^+$ as ``Model I" and the model with $N(2100)1/2^+$ as ``Model II". The corresponding parameters of these two models are listed in Table~\ref{table:constants}. Note that the cutoff value of $t$-channel $\rho$ and $\omega$ exchanges is taken from Ref.~\cite{Xing:2018axn}, where the value $\Lambda_M=1070$ MeV is determined by fitting the CLAS differential cross section data at $E_\gamma \approx 3.4$ GeV. We have checked that at this energy point the contributions from other interacting diagrams are negligible. This means that our results presented in this paper, even fitted to the LEPS2/BGOegg data at lower energies, can reproduce the CLAS data equally well as shown in Ref.~\cite{Xing:2018axn}. We don't repeat this plot here. 

We remark that adding a second or even more resonances in addition to either $N(2040)3/2^+$ or $N(2100)1/2^+$ to the model will definitely improve the theoretical description of data as more adjustable parameters will be introduced. Nevertheless, as the available data for $\gamma p \to f_0(980) p$ are rather limited, considering more resonances in the model will result in too many fits with similar $\chi^2/N$ and large error bars for the fitting parameters. One is then not able to draw any conclusive conclusion about the resonance contents in $\gamma p \to f_0(980) p$. Therefore, we postpone the analysis of adding two or more resonances to the model until more data for $\gamma p \to f_0(980) p$ become available in the future.

\begin{figure}[htb]
\includegraphics[width=\columnwidth]{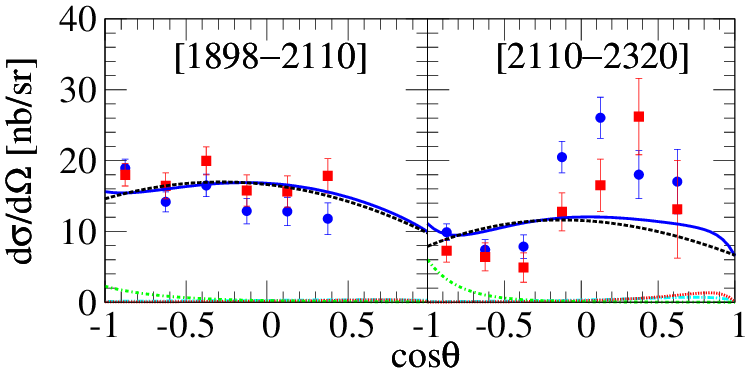}
\caption{Differential cross sections (blue solid lines) for $\gamma p \to f_0(980) p$ with $f_0(980)$ decaying into $\pi^0 \pi^0$ obtained from Model I. The cyan double-dot-dashed lines, red dotted lines, black dashed lines, and green dot-dashed lines denote the individual contributions from the $t$-channel $\rho$ exchange, $t$-channel $\omega$ exchange, $s$-channel resonance exchange, and $u$-channel $N$ exchange, respectively. The quantities in square brackets are the ranges of the center-of-mass energy of the system, in MeV. The data are the same as in Fig.~\ref{fig:xie}.}
\label{fig:section1}
\end{figure}

\begin{figure}[htb]
\includegraphics[width=\columnwidth]{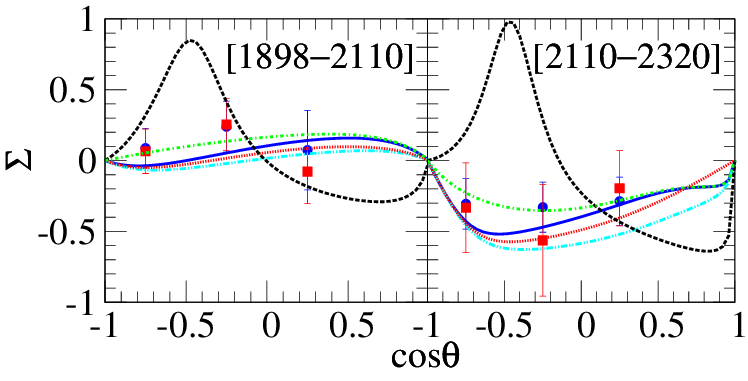}
\caption{Photon beam asymmetries (blue solid lines) $\Sigma$ for $\gamma p \to f_0(980) p$ obtained from Model I. The cyan double-dot-dashed lines, red dotted lines, black dashed lines, and green dot-dashed lines denote the results with the $t$-channel $\rho$ exchange, $t$-channel $\omega$ exchange, $s$-channel resonance exchange, and $u$-channel $N$ exchange being switched off, respectively. The quantities in square brackets are the ranges of the center-of-mass energy of the system, in MeV. The data are the same as in Fig.~\ref{fig:compare}.}
\label{fig:beam1}
\end{figure}

\begin{figure}[htb]
\includegraphics[width=\columnwidth]{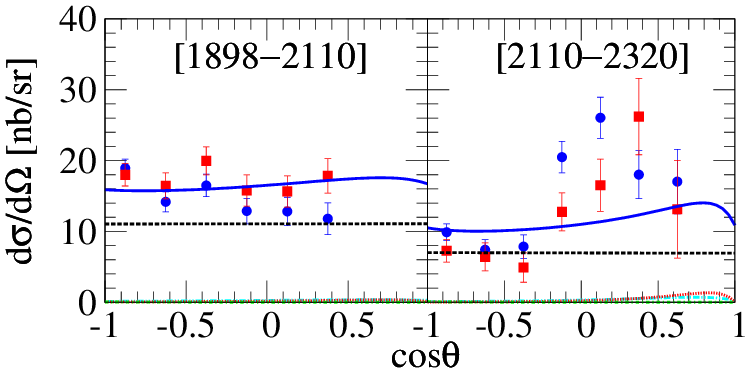}
\caption{Differential cross sections for $\gamma p \to f_0(980) p$ with $f_0(980)$ decaying into $\pi^0 \pi^0$ obtained from Model II. The quantities in square brackets are the ranges of the center-of-mass energy of the system, in MeV. Notations for lines and data are the same as in Fig.~\ref{fig:section1}.}
\label{fig:section2}
\end{figure}

\begin{figure}[htb]
\includegraphics[width=\columnwidth]{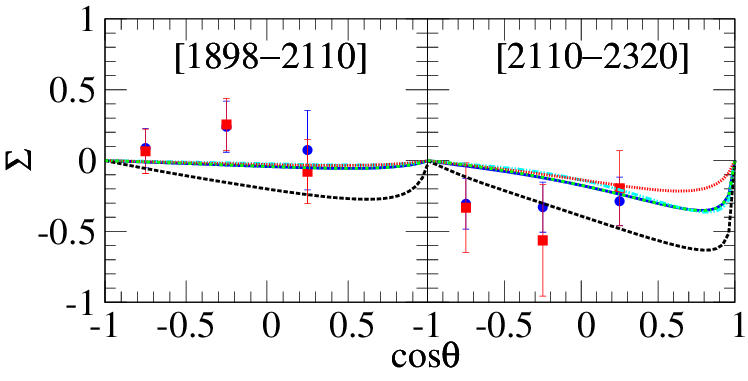}
\caption{Photon beam asymmetries for $\gamma p \to f_0(980) p$ obtained from Model II. The effects from individual contributions are calculated by switching off the corresponding amplitudes in the full reaction amplitudes. The quantities in square brackets are the ranges of the center-of-mass energy of the system, in MeV. Notations for lines and data are the same as in Fig.~\ref{fig:beam1}.}
\label{fig:beam2}
\end{figure}

\begin{figure}[htb]
\includegraphics[width=\columnwidth]{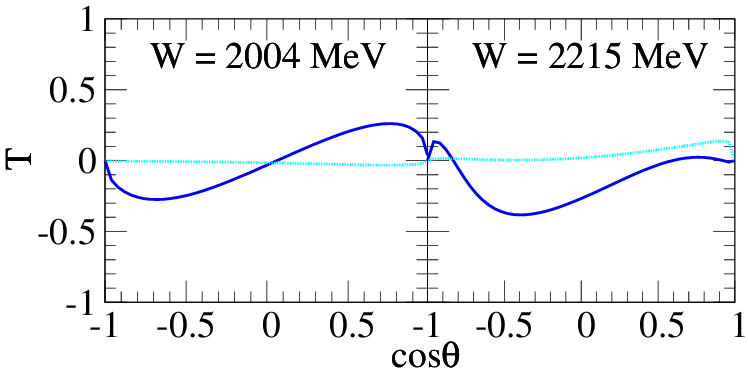}
\caption{Predictions of target asymmetries $T$ for $\gamma p \to f_0(980) p$ from Model I (blue solid lines) and Model II (cyan dotted lines).}
\label{fig:t}
\end{figure}

We now come to the detailed discussion of the results obtained from Model I. The differential cross sections and the photon beam asymmetries $\Sigma$ are shown in Fig.~\ref{fig:section1} and Fig.~\ref{fig:beam1}, respectively. We note that the LEPS2/BGOegg data \cite{LEPS2BGOegg:2023ssr} have been measured by two methods, with or without requirement of the $N^*$ cuts, and the difference between the two resulting data sets is modest compared with the evaluated statistical and systematic uncertainties \cite{LEPS2BGOegg:2023ssr}.

In Fig.~\ref{fig:section1}, the contributions from the individual terms of the $t$-channel $\rho$ and $\omega$ exchanges, the $s$-channel $N(2040)3/2^+$ exchange, and the $u$-channel $N$ exchange are also shown. The contributions from the $s$-channel $N$ exchange and the interaction current are too small to be clearly seen with the scale used and they are not plotted. We can see from Fig.~\ref{fig:section1} that the overall agreement of the model results with the data is reasonable. The contributions from the $s$-channel $N(2040)3/2^+$ exchange dominate the cross sections in the energy region of the LEPS2/BGOegg measurements. Noticeable contributions from the $t$-channel $\rho$ and $\omega$ exchanges can be seen at forward angles at the higher energy point. Besides, considerable contributions from the $u$-channel $N$ exchange can be seen at the backward angles.

For the beam asymmetries $\Sigma$ shown in Fig.~\ref{fig:beam1}, the effects of the individual terms of the $t$-channel $\rho$ and $\omega$ exchanges, the $s$-channel $N(2040)3/2^+$ exchange, and the $u$-channel $N$ exchange are calculated by switching off the corresponding amplitudes in the full reaction amplitudes. One can see that the full model results describe the data fairly well. The $s$-channel $N(2040)3/2^+$ exchange has very strong effect on beam asymmetries. The $t$-channel $\rho$ and $\omega$ have significant effects mainly at forward angles, and the $u$-channel $N$ exchange has significant effects mainly at backward angles. Effects from the $s$-channel $N$ exchange and the interaction current are very small and not shown in Fig.~\ref{fig:beam1}.

The results obtained from Model II are shown in Fig.~\ref{fig:section2} and Fig.~\ref{fig:beam2} for the differential cross sections and beam asymmetries $\Sigma$, respectively. Overall, the agreement of the model results with the data is also reasonable. In Fig.~\ref{fig:section2}, one can see that the contributions from the $s$-channel $N(2100)1/2^+$ exchange dominate the cross sections of this reaction. As expected, the contributions from the $t$-channel $\rho$ and $\omega$ exchanges are the same as the ones in Model I. However, unlike the situation in Model I, the contributions from the $u$-channel $N$ exchange are now much smaller in Model II. This is simply because the fitted cutoff values $\Lambda_B = 958$ and coupling constant $|g_{NNf_0}| = 2.08$ obtained in Model I are much larger than the corresponding values $\Lambda_B = 800$ MeV and $|g_{NNf_0}| = 0.65$ obtained in Model II, as shown in Table~\ref{table:constants}. The contributions from the $s$-channel $N$ exchange and the interaction current are too small and they are not shown in Figs.~\ref{fig:section2}. For the beam asymmetries $\Sigma$ shown in Fig.~\ref{fig:beam2}, we can see that the $s$-channel $N(2100)1/2^+$ exchange has relatively stronger effects, and the $t$-channel $\omega$ exchange has considerable effects at forward angles at the higher energy point.

In Fig.~\ref{fig:t}, the predictions of the target asymmetries $T$ at $W = 2004$ MeV and $W = 2215$ MeV for $\gamma p \to f_0(980) p$ obtained from Model I and Model II are shown. One sees the predicted target asymmetries from these two models are quite different. Data from future experiments on this observable are called on to distinguish Model I and Model II for the $\gamma p \to f_0(980) p$ reaction.

In brief, by including the contributions from either the $N(2040)3/2^+$ or $N(2100)1/2^+$ resonances, the data from the LEPS2/BGOegg Collaboration \cite{LEPS2BGOegg:2023ssr} on both the differential cross sections and the photon beam asymmetries for $\gamma p \to f_0(980) p$ can be well reproduced. In both cases, contributions from individual reaction mechanism are discussed and  predictions of the beam-target asymmetry are given to examine the models.

\section{Summary and conclusion}  \label{sec:summary}

The data from the LEPS2/BGOegg Collaboration \cite{LEPS2BGOegg:2023ssr} on differential cross sections and photon beam asymmetries for the $\gamma p \to f_0(980) p$ reaction are analyzed within a tree-level effective Lagrangian approach. The theoretical model is constructed by considering the contributions from the $t$-channel $\rho$ and $\omega$ exchanges, the $u$- and $s$-channel $N$ exchanges, the interaction current, and the $s$-channel $N(2040)3/2^+$ or $N(2100)1/2^+$ resonance exchange mechanisms. The contributions from the $s$-channel $N(2040)3/2^+$ or $N(2100)1/2^+$ resonance exchange are found to dominate the differential cross sections and have significant effects on the photon beam asymmetries in the energy region considered. The contributions from the $u$-channel $N$ exchange are considerable in the case of including $N(2040)3/2^+$ while negligible in the case of including $N(2100)1/2^-$. Predictions of the target asymmetries $T$ for $\gamma p \to f_0(980) p$ in both models with either $N(2040)3/2^+$ or $N(2100)1/2^-$ resonance are given, which is expected to be examined by the future experiments.

\begin{acknowledgments}
The author Neng-Chang Wei thanks N. Muramatsu for kindly providing the experimental data for the present work. This work is partially supported by the National Natural Science Foundation of China under Grants No.~12305137, No.~12175240, and No.~11635009, the Fundamental Research Funds for the Central Universities, and the China Postdoctoral Science Foundation under Grants No.~2021M693141 and No.~2021M693142.
\end{acknowledgments}

\end{document}